\documentclass[aps,prl,twocolumn,groupedaddress,showpacs]{revtex4}

\usepackage{graphicx}% Include figure files
\usepackage{dcolumn}% Align table columns on decimal point
\usepackage{bm}% bold math
\usepackage{epstopdf}
\usepackage{amsmath}
\usepackage{color}

\begin{document}

\title{Non-equilibrium probing of two-level charge fluctuators using the step response of a single electron transistor}
\author{A.~Pourkabirian$^1$, M.V.~Gustafsson$^{1}$, G.~Johansson$^1$, J.~Clarke$^{1,2}$, P.~Delsing$^1$}
\affiliation{$^1$Microtechnology and Nanoscience, Chalmers University of Technology, S-41296, G\"{o}teborg, Sweden}
\affiliation{$^2$Department of Physics, University of California, Berkeley, California 94720, USA.}
\date{\today}

\begin{abstract}

We report a new method to study two level fluctuators (TLFs) by measuring the offset charge induced after applying a sudden step voltage to the gate electrode of a single electron transistor. The offset charge is measured for more than 20 hours for samples made on three different substrates. We find that the offset charge drift follows a logarithmic increase over four orders of magnitude in time and that the logarithmic slope increases linearly with the step voltage. The charge drift is independent of temperature, ruling out thermally activated TLFs and demonstrating that the charge fluctuations involve tunneling. These observations are in agreement with expectations for an ensemble of TLFs driven out of equilibrium. From our model, we extract the density of TLFs assuming either a volume density or a surface density.

\end{abstract}

\maketitle

Two level fluctuators (TLFs) are found in many, if not all, solid state systems. The microscopic origin and physics of TLFs have been extensively studied in mesoscopic physics over the last 3 decades \cite{Paladino2014,Simkins2009,Rees2008,KenyonIEEE1999,FurlanPRB2003,ZimmermanPRB1997,ZimmermanSiOffset}. It is generally believed that an ensemble of TLFs gives rise to charge noise with a $1/f$-like power spectrum which limits the performance of all charge sensitive devices \cite{RogersBuhrmanPRL,Machlup1954}, including single electron transistors (SETs) \cite{SETLikharev}, quantum point contacts \cite{Qpointcontact} and quantum capacitance detectors \cite{Qcapacitance}. TLFs also induce decoherence in solid state qubits that are the building blocks in quantum information processing \cite{ClarkeWilhelm,Bylander2011,MakhlinRevModPhys}. Although there are various models for TLFs \cite{KafanovPRB2008, MG_PRB, DuBoisPRL2013,kenyonTSq}, their physical origin and location remain open questions. The simplest microscopic model consists of a two-well potential containing a charged particle. Depending on the height and width of the barrier separating the wells and on temperature, the particle is transferred from one well to the other either by thermally activated hopping \cite{DuttaHorn,PRLNECTsq} or by tunneling \cite{KimPRB}. There are also different scenarios regarding the location of the TLFs: they may be distributed in the volume of the substrate (volume distribution) \cite{ZorinPRB1996}, or at the interfaces between metals and insulators (surface distribution).
Earlier we suggested \cite{MG_PRB} that charge noise may arise from electrons tunneling back and forth between the Fermi sea in the metallic electrode and traps at the metal-insulator interface such as localized metal induced gap states (MIGs); MIGs have been proposed as a possible origin of the localized magnetic moments giving rise to flux noise in SQUIDs and flux qubits \cite{ChoiPRL2009}.

SETs are used as electrometers to study TLFs. The SET [Fig.\,1(a)] is extremely sensitive to charge \cite{Aassime2001ultrasen, Brenning2006ultrasen}. When the SET is voltage biased, the current, $I_{SET}$, is periodically modulated by the charge induced on its island by a nearby gate [Fig.\,1(a)] with period $e/C_{g}$. Here, $C_{g}$ is the capacitance between the island and the gate electrode, and \textit{e} is the electron charge. Consequently, a fluctuating charge in the vicinity of the SET causes $I_{SET}$ to fluctuate.

\begin{figure}
\includegraphics[width = 1\columnwidth]{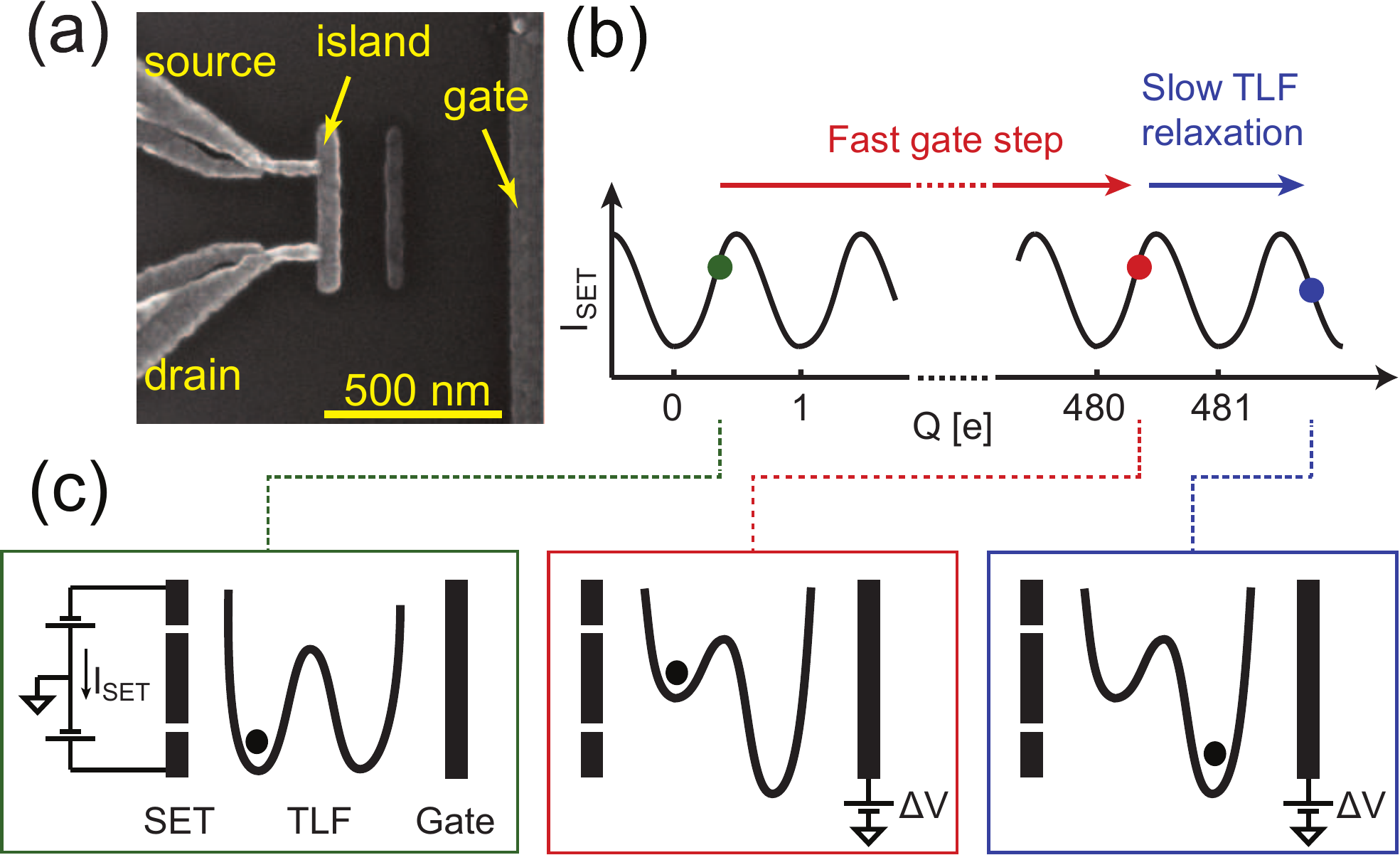}
\caption{(Color online) (a) Scanning electron micrograph of a SET with typical junction size of 20$\times$50\,nm$^2$. (b,c) Schematic overview of the measurements. Starting from equilibrium (green), we apply a high and sudden voltage step $\Delta V$ to the gate, bringing the TLF ensemble out of equilibrium (red). After the step, we use the SET to record the charge drift $Q$ over time as the TLFs relax to their new ground states (blue). }
\label{Fig1}
\end{figure}

In virtually all previous work on TLFs and charge noise, the ensemble of TLFs is close to equilibrium and the data acquired in these experiments have generally not conclusively shown whether the mechanism is thermal activation or tunneling.

In this Letter, we take another approach and investigate the response of TLFs driven far out of equilibrium by a strong external electric field. We present measurements of the charge drift, $Q$, following the application of a sudden step voltage, $\Delta V$, to the SET gate, which causes the induced charge on the SET island to increase rapidly. Remarkably, the charge drifts slowly long after the step is applied [Fig.\,1(b)]. We argue that this drift is due to the change in the potential landscapes of the TLFs caused by the gate voltage. Some of the TLFs are brought to metastable states, which decay after a characteristic time causing the charge drift [Fig.\,1(c)]. We have measured this drift for up to 20 hours for several SETs made on three different kinds of substrates. We find that the drift increases logarithmically with time and is independent of temperature, allowing us to rule out thermally activated TLFs. Furthermore, we have measured the response to voltage steps with different heights, and found that the logarithmic slope of the drift increases with increasing voltage. We show that this behavior is consistent with the response of an ensemble of TLFs, and we develop a theory from which we can extract the densities for these TLFs assuming either volume or surface distributions.

\begin{table}[b]
 %\centering
\caption{\label{tab:table1}%
Extracted parameters for the four measured SETs. }
\begin{ruledtabular}
\begin{tabular}{lccccc}
Device&Substrate&$E_C/k_{B}$\footnote{$E_{C}$ is the charging energy of the SET, the energy required to charge the SET island with one electron.}&$C_{g}$&$T$&$H=\frac{1}{\Delta V}\frac{\Delta Q(t)}{\Delta \textrm{log}\,t}$\\
 & &(K)&(aF)&(mK)&[$e/(\textrm V \cdot \textrm{decade of } t$)] \\

\hline
1&Si-SiO$_{X}$&4.1&7.8&2000&0.26 \\
1&Si-SiO$_{X}$&4.1&7.8&30&0.22 \\
2&Si-SiO$_{X}$&5.8&10&20&0.22 \\
3&Glass&4.2&9.2&20&0.37 \\
4&Sapphire&2.3&18.5&20&0.19 \\
\end{tabular}
\end{ruledtabular}
\end{table}

We made measurements on samples with nominally identical layouts fabricated on three different substrates: glass, sapphire and oxidized silicon with an oxide thickness of 400\,nm. The aluminum SETs were fabricated with electron-beam lithography and double-angle evaporation \cite{Dolan1977}. All measurements were performed in a dilution refrigerator with a base temperature of 20\,mK. A magnetic field of 1\,T quenched superconductivity in the aluminium. We present results for four representative samples, with extracted parameters shown in Table I.

\begin{figure}[t]
\includegraphics[width = 1\columnwidth]{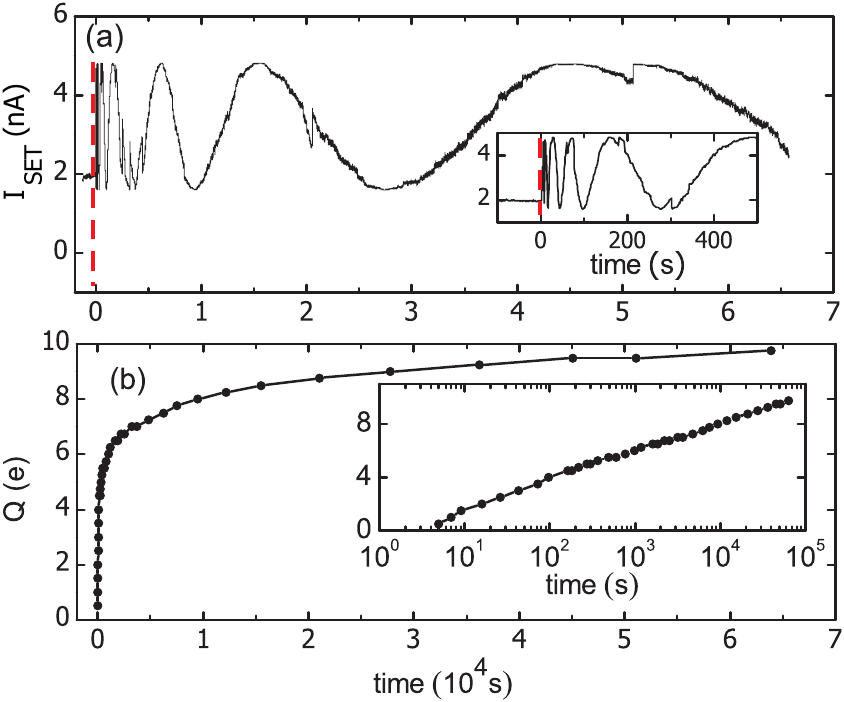}
\caption{(Color online) Step response measurements at $T= 30\,$mK on device\,1 on an oxidized silicon substrate. (a) Continuous measurement of $I_{SET}$ over a period of about 20 hours. A step voltage $\Delta V=9.8$\,V, was applied to the gate at $t=0$ shown as red dashed lines. The inset shows the first 500 seconds after the step. (b) Charge drift extracted from (a). The inset shows the same data on a logarithmic time axis; the charge increases logarithmically with time.}
\label{Fig2}
\end{figure}

In all experiments, the SET was biased symmetrically with respect to ground and we stepped the gate voltage abruptly while sampling $I_{SET}$ continuously at 2\,ksamples/s. Figure 2(a) shows $I_{SET}\ \textit{vs.}$ time for a SET made on an oxidized silicon substrate (device\,1) at a temperature $T=30\,$mK. The sample was left for a long time at a gate voltage $V_g=-4.9\,$V and $V_g$ then stepped to +4.9\,V, giving a step height of $\Delta V=9.8\,$V. Figure 2(b) shows the charge induced on the island extracted from the data in (a) by counting the number of oscillations in $I_{SET}$, each one of which corresponds to an additional electron induced on the island. The inset in Fig.\,2(b) shows the same data on a logarithmic scale. We see that the offset charge increases logarithmically over more than four decades of time.
We note that we cannot count the total number of electrons induced on the island since hundreds of electrons are induced instantaneously when we apply the voltage step. We can, however, estimate the initial change in charge from the gate capacitance $C_{g}$. For device\,1 the initial charge jump was approximately 480\,$e$. Thus, the additional slow drift of about 10\,$e$ follows the initial step of 480\,$e$ [Fig.\,1(b)].

To obtain a direct measurement of the charge drift, we used a proportional-integral-derivative (PID) regulator in subsequent experiments, feeding the regulation signal to the gate to maintain a constant SET current. Since the regulation commenced $\sim$1\,s after the voltage step was applied, we cannot capture the first few seconds of the drift.
Figure 3 shows the measured charge drift using both methods for different devices for almost 20 hours after applying the voltage step. From the data in Fig.\,3, we extract the logarithmic slopes of the charge drift normalized to the step height,
\begin{equation}
 H=\frac{1}{\Delta V}\frac{\Delta Q(t)}{\Delta \textrm{log}\,t},
\end{equation}

\noindent summarized in Table I. Comparing the two measurements for device\,1 at 30\,mK and 2\,K we see that $H$ does not depend significantly on $T$ in this range.

\begin{figure}[t]
\includegraphics[width=1\columnwidth]{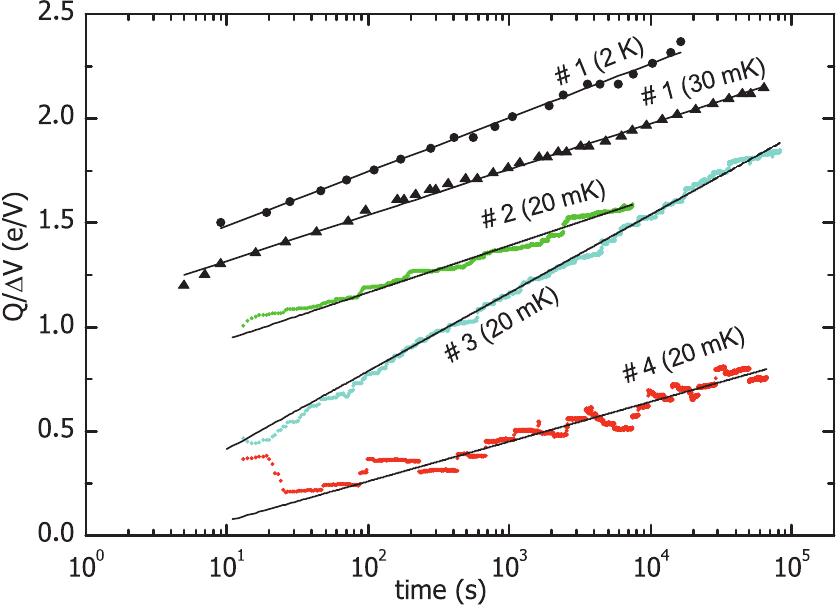}
\caption{(Color online) Normalized charge drift measured for four different devices. Plots have been offset vertically for clarity. For device\,1 the charge drift was extracted from the current modulation of the SET. For devices 2, 3, and 4 a PID loop was connected to the gate (see text) and the data were sampled at 2 ksamples/s to improve the resolution of the charge drift. The measurements on different SETs on different substrates show a logarithmic charge drift with similar slopes. The curves have been shifted vertically for clarity. The charge drift per decade of time in Table I is extracted from a least squares fit to each trace (solid black lines).}
\label{Fig3}
\end{figure}

To investigate the dependence of the total measured drift on step height, in separate measurements we applied voltage steps with different heights to device\,3 (glass substrate) and measured the charge drift. Figure 4 shows that the rate of charge drift is proportional to $\Delta V$.

The simplest microscopic model for a TLF is a charged particle trapped in a double-well potential with an energy difference $\Delta E=E_R-E_L$ between the right and the left well, and with an energy barrier of height $E_b$ [Fig.\,5(a,b)]. The charge, which we assume to be the electron charge $e$, moves a distance $d$ between the two locations either by thermal activation over the barrier [Fig.\,5(a)] or by tunneling through the barrier [Fig.\,5(b)]. The motion is characterized by a switching time $\tau= 1/\omega_0$, where $\omega_0$ is the sum of the forward and backward rates which depend on the properties of the TLF potential and on $T$. The equilibrium population of the right well is given by the Fermi distribution $f(\Delta E)$.
An alternative model is a charge that moves between the Fermi gas in one of the electrodes and a well with energy $\Delta E$ compared to the Fermi energy and energy barrier $E_b$. For the purpose of this work these two models behave in the same way, but we base our description on the double well TLF.

To illustrate how the TLFs influence the SET in our experiment, we consider a simple parallel-plate capacitor model, where one plate consists of the gate and the other one the SET island and leads \cite{Geometry}. Figure 5(c) shows such a geometry with a single TLF, with angle $\theta$ between its displacement vector, $\vec{d}$, and the gate field lines, $\vec E_{G}(\vec r)$ [solid lines in Fig.\,5(c)].

First, we consider the effect on the TLFs of a voltage step applied to the gate. The electric field changes the energy difference between the two wells  by $\delta E=e \Delta V \vec{d} \cdot \vec e_{G}(\vec r) $, where we have defined the normalized gate field $\vec e_G(\vec r) = \vec E_{G}(\vec r)/\Delta V$. To determine how much charge a TLF in a given location induces on the SET island, we calculate the electric field $\vec{E}_V(\vec r)$ in a \emph{virtual} situation, where the island is held at a potential $V_{0}$ and the gate, source, and drain are grounded [dashed lines in Fig.\,5(c)]. We define the normalized virtual field as $\vec e_V(\vec r)=\vec E_{V}(\vec r)/V_0 $. For a TLF at any point $\vec r$ in space, the change in charge induced on the island by a switching event is given by
$\delta q(\vec r) = e\,\vec{d} \cdot \vec{e}_V(\vec r)$ \cite{Smythe,Mathieson}.

The equilibrium population of each TLF is determined by its energy difference $\Delta E$, which suddenly changes by an amount $\delta E$ as the gate voltage step is applied [Fig.\,5(a,b)]. The charge distribution approaches the new equilibrium on a timescale set by the (new) switching time $\tau'$. We assume an ensemble of TLFs with different final switching times, $\tau'$, different initial energy differences, $\Delta E$, and different displacement vectors, $\vec{d}$ and sum the contribution from a set of \textit{N} TLFs to find the total charge induced on the island
\begin{equation}
Q(t)=\sum_{i=1}^N \delta q_i \left[f(\Delta E_i)-f(\Delta E_i-\delta E_i)\right](1-e^{-t/\tau_i'}) ,
\end{equation}

\noindent where the subscript $i$ refers to the individual TLF values.

\begin{figure}
\includegraphics[width=1\columnwidth]{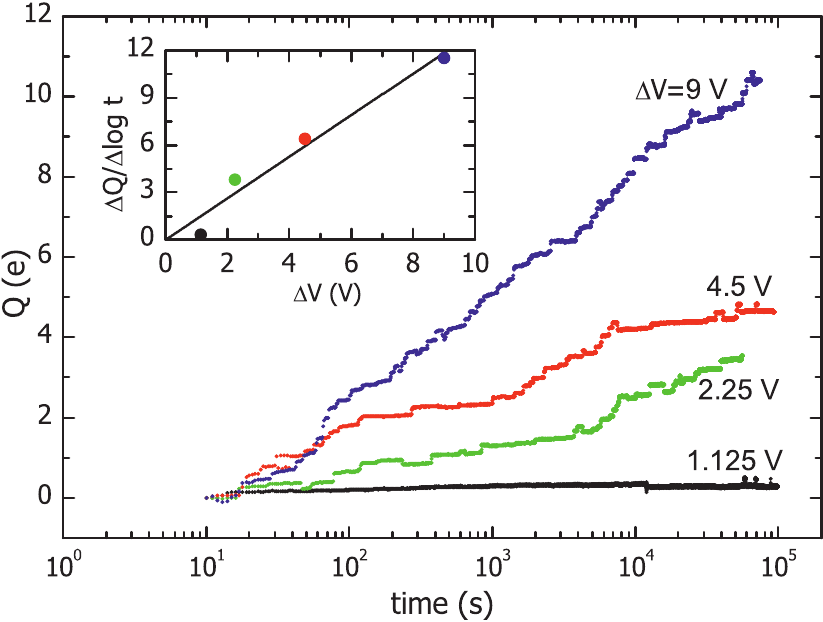}
\caption{(Color online) Charge drift for different voltage step heights, $\Delta V$, for device\,3 (glass substrate) at 20\,mK. The logarithmic slope of the charge drift, $\Delta Q(t)/\Delta$log$\,t$, increases with the voltage step amplitude. The inset shows $\Delta Q(t)/\Delta$log$\,t$ \textit{vs.} $\Delta V$, and the line represents a least squares fit constrained to pass through the origin. }
\label{Fig4}
\end{figure}

We now assume (\textit{i}) that the TLFs are numerous enough to change the sum into an integral, (\textit{ii}) a flat distribution of initial energy differences, (\textit{iii}) a logarithmic distribution of switching times $\tau$, between minimum and maximum switching times $\tau_{\rm min} = 1/\omega_{\rm max}$ and $\tau_{\rm max} = 1/\omega_{\rm min}$. These assumptions lead to the observed $1/f$ power spectrum for the noise \cite{MG_PRB} and would arise naturally from a roughly flat distribution of barrier heights in the case of thermal activation, and barrier widths in the case of quantum tunneling when the barrier height is the largest energy of the system. For $\tau _{\min } \ll t \ll \tau _{\max }$ we find \cite{Supplemental}
\begin{equation}
Q(t) \approx \frac{\ln (\omega_{max}t)+\gamma }{{\ln (10)}}\int {n(\vec{r},\theta)\,\delta q(\vec{r},\theta)\,\delta E(\vec{r},\theta) \,\mathrm{d}\vec{r}\mathrm{d}\theta},
\end{equation}

\noindent where $\gamma$ is Euler's gamma and $n(\vec{r},\theta)$ is the density of TLFs with displacement vector oriented along the direction $\theta$, per unit energy difference and per decade of switching time.
In the case of {\em quantum tunneling} the switching rates are independent of temperature, and the effect of temperature is basically to broaden the population according to the Fermi distribution. This has no effect on the step response.
In the case of {\em thermal activation}, the switching rate is determined by $T, E_b$, and the attempt frequency $\Omega$ according to $\omega_0=\Omega e^{-E_b/k_BT}$. If we consider $\Omega$ and $E_b$ to be independent of $T$ and that $E_b$ has a flat distribution, the TLF density per decade scales linearly with $T$. It follows from Eq.\,(3) that the step response should then also be proportional to $T$ \cite{Supplemental}.
In our experiments $\delta E \ll k_BT$, and therefore the voltage step actually flattens out the TLFs which are thermally active at equilibrium, since $E_b \sim k_B T$. These TLFs will thus switch immediately, and only those with a remaining barrier height comparable to temperature will contribute to the slow charge drift. Assuming a flat distribution of final barrier heights, the slow charge drift would actually be independent of $\Delta V$.

\begin{figure}
\includegraphics[width = .95\columnwidth]{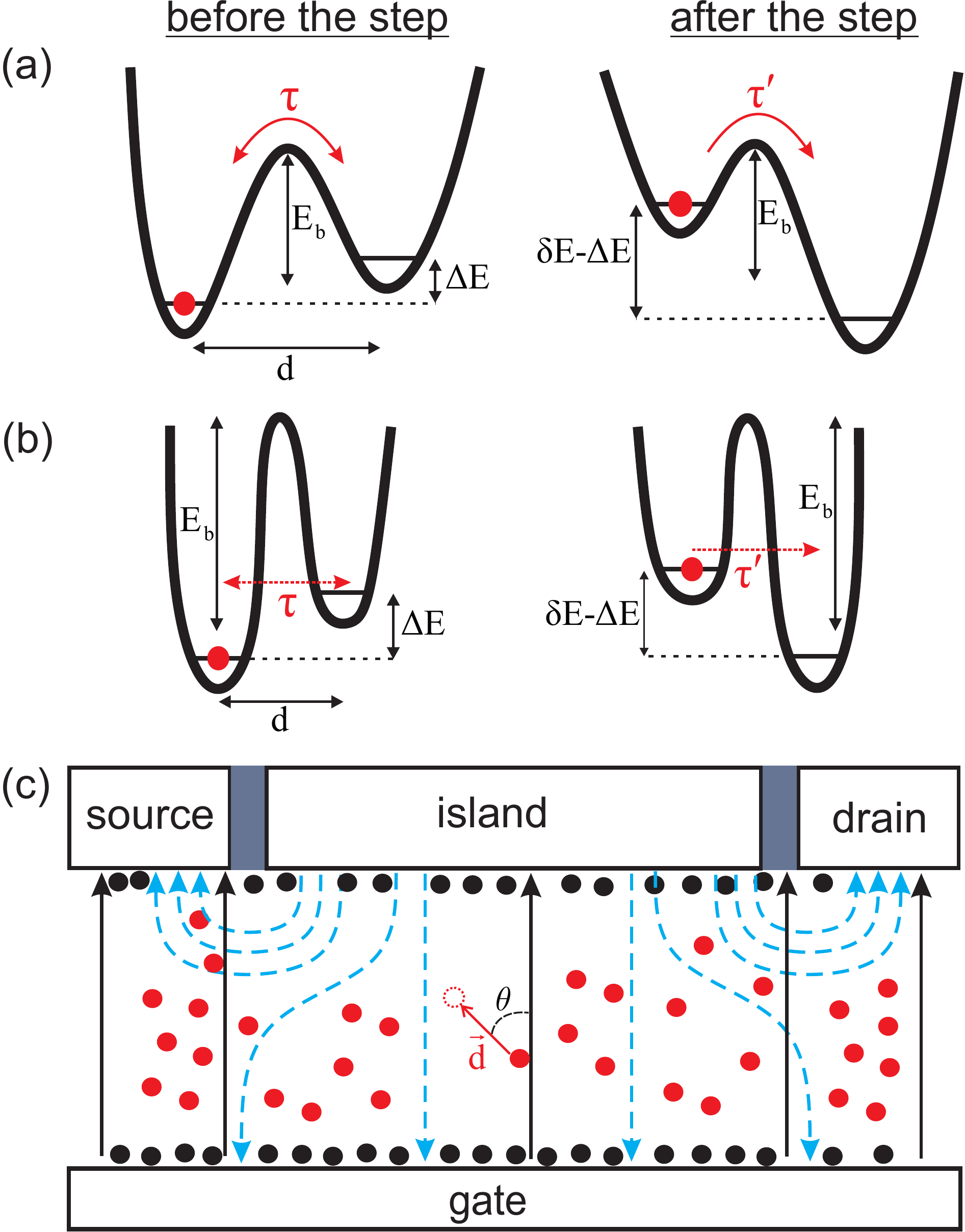}
\caption{Microscopic model of a TLF and its influence on the SET. (a,b) A charged particle in a double potential well with a distance $d$ between the two wells, barrier height $E_{b}$, and energy difference $\Delta E$ between the two states. Left: before applying the step, the switching time is $\tau$. Right: after applying the step, the energy difference between the two wells changes by $\delta E$, and the switching time changes to $\tau'$. $E_b$ is defined with respect to the mean between the two states and does not change to first order. (a) Thermal activation. (b) Quantum tunneling.
(c) Simplified geometry of the SET, gate and TLFs. At the center, an individual TLF is shown schematically with an angle $\theta$ between its displacement vector $\vec{d}$ and the gate field $\vec E_{G}(\vec r)$ (solid lines). The dashed lines show the \emph{virtual} field, $\vec E_{V}(\vec r)$ (see text). The red and black circles represent the volume and surface distributions of TLFs, respectively.}
\label{Fig5}
\end{figure}

To compare with experiment we calculate the parameter $H$, the induced charge per decade of time and per applied gate voltage [Eq.\,(1)]. We consider two special cases: homogeneous volume and homogeneous surface distributions of $n(\vec{r})$. In the case of a homogeneous volume distribution, $n_{v}$, we assume that the TLF can be randomly oriented. We obtain \cite{Supplemental}
\begin{equation}
H_v=\frac{e^2\,d^2\,n_v}{3} \int \vec{e}_V(\vec r)\cdot \vec{e}_G(\vec r) \, \mathrm{d}V.
\end{equation}

\noindent For a homogeneous surface distribution, $n_{s}$, we assume that the electrons tunnel perpendicularly from the metal surface, $S$, to the trap, \textit{i.e.} $\theta=0$. We obtain
\begin{equation}
H_s=e^2\,d^2\,n_s \int \vec{e}_V(\vec r)\cdot \vec{e}_G(\vec r) \, \mathrm{d}S.
\end{equation}

\noindent It is interesting to note that the change in charge can be either positive or negative. In particular, when a positive voltage is applied to the gate, a TLF situated directly underneath the SET island and a TLF underneath the drain or source will both switch in the same direction. However, since $\vec e_{V}(\vec r)$ points in opposite directions at the two locations, the \emph{induced charge} from the TLFs underneath the drain and source will have the opposite sign compared to the contribution from the TLFs underneath the SET island \cite{Supplemental}. Thus, the sign of a particular charge change (see \textit{e.g.} device\,4 in Fig.\,3) provides information about the location of an individual TLF.

Our experimental data clearly show that the charge on the SET increases logarithmically with time after a voltage step has been applied to the gate. Although charge drift with a similar behavior has been reported previously \cite{Zimmerli1992,Manninen2001,Savin2001a}, the dependence on time was not analyzed and in Ref.\,\cite{Zimmerli1992} the charge drift was discussed in terms of a leakage resistance. For a leakage resistance, however, one would expect the charge to increase linearly with time, contrary to our observations.

Studying the details of this charge drift we draw a number of conclusions about the TLFs. (i) The charge drift appears not to depend on temperature. Measurements on device\,1 at both 30\,mK and 2\,K show very similar values for $H$ (Table I), indicating that the charge transfer mechanism is tunneling and not thermal activation: If the TLFs were thermally activated we would expect $H$ to depend on temperature. (ii) We find that the charge drift increases approximately linearly with $\Delta V$ (Fig.\,4), indicating that the distribution of $\Delta E$ for the TLFs is uniform. This also speaks against thermal activation, since in that case, the drift would not depend on $\Delta V$ \cite{Supplemental}. Furthermore, tunneling is consistent with our previous measurements of linear temperature dependence of the charge noise spectral density \cite{MG_PRB}. (iii) The charge drift is similar for devices fabricated on different materials. (iv) Using the model described earlier and the measured values for $H$, we extract the density of TLFs from Eqs.\,(4,5) by calculating the integrals numerically for the actual geometry \cite{Supplemental}. Assuming $d\approx 1$ nm, we estimate the densities to be $ n_{v}\approx 1.5 \times10^{24}$ \textrm{(m$^{3}$ eV decade)$^{-1}$} and $n_{s}\approx 1.6 \times10^{16}$ \textrm{(m$^{2}$ eV decade)$^{-1}$}, for the volume and surface cases respectively. The extracted surface density is similar to the density predicted for MIGs \cite{ChoiPRL2009}.

The work was supported by the Swedish Research council and by the Wallenberg Foundation. We thank SangKook Choi, Jared Cole, Steven Louie, Samira Nik, Eva Olsson and Lunjie Zeng for very helpful discussions. JC thanks the faculty of the Chalmers University for their hospitality during his tenure of the Chalmers 150th Anniversary Visiting Professorship during which some of the experiments were performed. His research was funded by the Office of the Director of National Intelligence (ODNI), Intelligence Advanced Research Projects Activity (IARPA), through the Army Research Office.

\bibstyle{apsrev}
\bibliography{References}

\begin{thebibliography}{34}
\bibitem[{Geo()}]{Geometry}
\bibinfo{note}{The calculations however are performed using the real geometry
  of the device.}
\end{thebibliography}

\end{document}

\title{Non-equilibrium probing of two-level charge fluctuators using the step response of a single electron transistor: Supplemental Material}
\author{A.~Pourkabirian$^1$, M.V.~Gustafsson$^{1}$, G.~Johansson$^1$, J.~Clarke$^{1,2}$, P.~Delsing$^1$}
\affiliation{$^1$Microtechnology and Nanoscience, Chalmers University of Technology, S-41296, G\"{o}teborg, Sweden.}
\affiliation{$^2$Department of Physics, University of California, Berkeley, California 94720, USA.}
\date{\today}

\maketitle

\section{I. Derivation of TLF densities }

\subsection{TLF energy shift induced by the gate voltage step}

\noindent When a voltage is applied to the gate of the SET, an electric field, $\vec{E}_G(\vec r)$, is generated between the gate and the island, source and drain collectively. We refer to this field as the gate field and note that it varies in space and depends on the geometry. We can neglect the potential of the island and treat all the three leads as grounded since the applied step voltage on the gate, $\Delta V$, is much larger than the Coulomb blockade voltage, $e/C_{\Sigma}$, where $C_{\Sigma}$ is the total capacitance between the island and all other leads. Since the gate field grows linearly with $\Delta V$ we can also define the normalized gate field, which has the units of inverse length, as

\begin{equation}
\vec{e}_G(\vec r)=\frac{\vec{E}_G(\vec r)}{\Delta V}.
\end{equation}

\noindent The displacement vector, $\vec{d}$, defines the distance, $d$, and the direction along which the charge of the TLF moves (either by thermal activation or by quantum tunneling) with respect to gate field direction. The voltage step shifts the double well potential of the TLF by

\begin{equation}
\label{dE}
\delta E(\vec{r},\vec{d}) = e \vec{d} \cdot \vec{E}_G(\vec r) =  e d \Delta V e_G(\vec r) \cos(\theta),
\end{equation}

\noindent where $e_G(\vec r)$ is the absolute value of $\vec{e}_G(\vec r)$ and $\theta$ is the angle between $\vec{e}_G(\vec r)$ and $\vec{d}$. For subsequent calculations, we assume that the moving charge is the electron charge, \textit{e}, and that the distance \textit{d} is of the order of 1 nm.

\begin{figure}[h]
\includegraphics[width = 0.4\columnwidth]{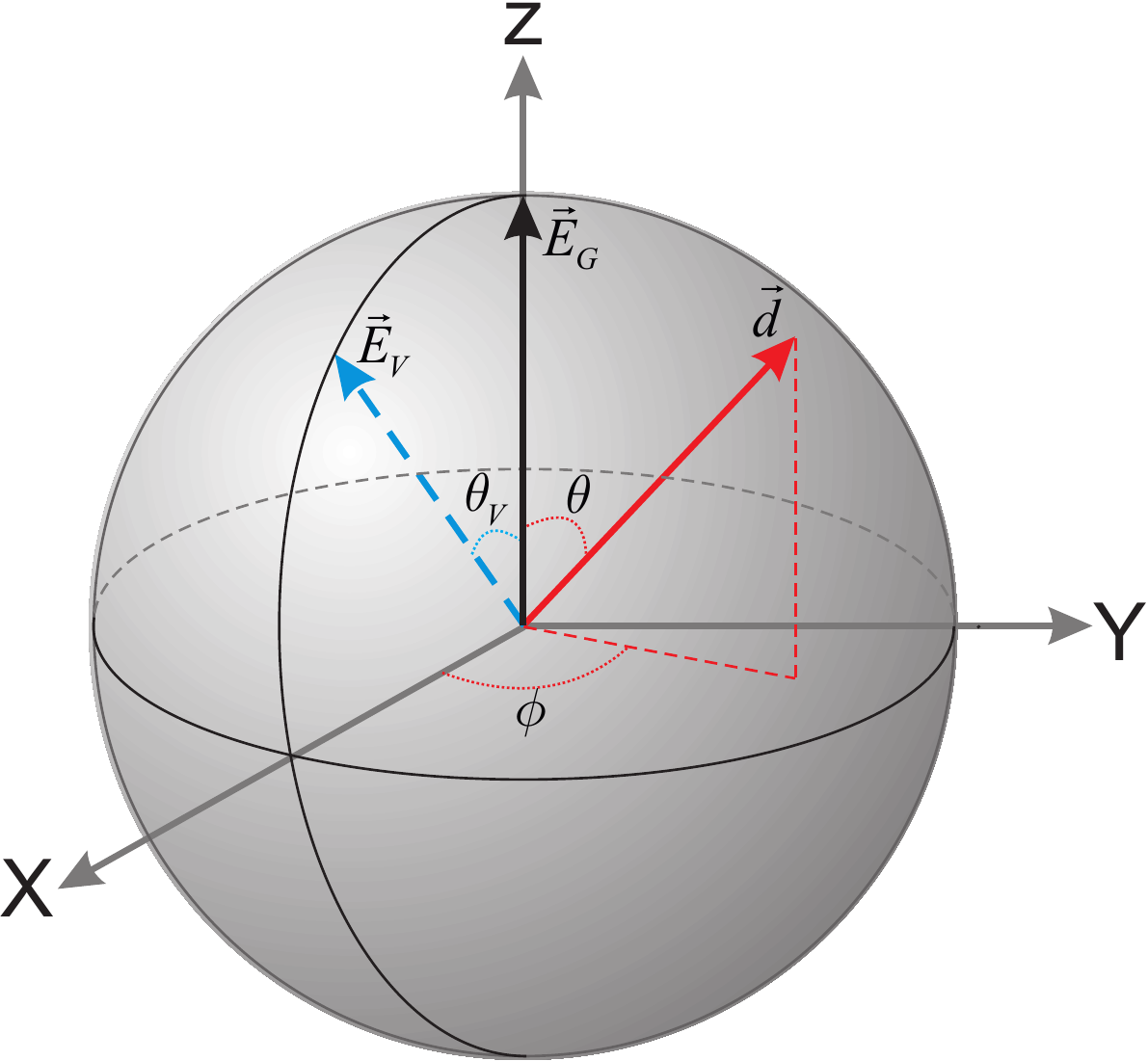}
\caption{The coordinate system and fields at the location of a TLF. Without loss of generality, we orient our coordinate system such that the gate field is aligned with the z-axis (black arrow). Furthermore, the coordinate system is aligned so that the virtual field is in the xz-plane (blue arrow). The red arrow represents the displacement vector $\vec{d}$.}
\label{FigS1}
\end{figure}

\subsection{Charge induced on the SET island by a switching TLF}

\noindent The next step is to calculate the charge induced on the SET island when a charged particle in the TLF moves from one well to the other . To do so, we define the virtual field, $\vec{E}_V(\vec r)$, which is the field we obtain by applying the potential $V_0$ to the SET island with all other electrodes, gate, source and drain, grounded. Again we define a normalized virtual field as $\vec{e}_V(\vec r) = \vec{E}_V(\vec r) / V_0$ which also has units of inverse length.
The charge induced on the SET island by a particular TLF, $\delta q$, is given by the scalar product between the displacement vector for that TLF, $\vec{d}_i$, and
the virtual field vector at the location of the TLF,

\begin{equation}
\label{dq}
\delta q(\vec{r},\vec{d})= e \vec{d} \cdot \vec{e}_{V}(\vec{r},\vec{d}).
\end{equation}

\noindent Figure\,S1 shows the coordinate system and the fields at the location of the TLF. Without loss of generality, we orient our coordinate system such that the gate field is aligned with the z-axis (black arrow in Fig.\,S1). Furthermore, the coordinate system is aligned so that the virtual field is in the xz-plane (blue arrow). The red arrow represents the displacement vector $\vec{d}$. We find

\begin{equation}
\vec{d}=d \left[\sin \left(\theta \right) \cos \left(\phi _d\right),\sin \left(\theta \right) \sin \left(\phi _d\right),\cos \left(\theta \right)\right]\\
\end{equation}

\noindent and

\begin{equation}
\vec{e}_V(\vec r)=e_V(\vec r)\left[\sin \left(\theta _V\right),0,\cos \left(\theta _V\right)\right].
\end{equation}

\noindent We now express $\delta q(\vec r, \vec d)$ in terms of $d$, $e_V(\vec r)$, and the angles between the two fields :

\begin{equation}
\delta q(\vec{r},\vec{d})=e\vec{d}\cdot \vec{e}_V(\vec r)=e\ d\ e_V(\vec r) \left[\cos \left(\phi \right)\sin \left(\theta \right) \sin \left(\theta _V\right)+\cos
\left(\theta \right)\cos \left(\theta _V\right)\right].
\end{equation}

\subsection{Which TLFs are activated by the gate voltage step?}

The equilibrium population of the excited state of the TLF is given by the Fermi function $f(\Delta E)$, where $\Delta E$ is either the energy of the single well compared to the Fermi energy, or the (positive) energy difference between the two wells of the double well. This is generally true when the excitation rate $\Gamma_{\rm exc}$ is thermally suppressed compared to the relaxation rate according to $\Gamma_{\rm rel}=e^{-\Delta E/k_BT}\Gamma_{\rm exc}$. This is the case in the double well potential, both for thermal excitation considering equal attempt rates for the two wells, as well as for phonon- or photon-assisted tunnelling between the two wells, considering a phonon- or photon-bath in thermal equilibrium. For the single well, the population is trivially given by the Fermi function when the spin lifetime of the electron is short. In the case of an infinite spin lifetime, both spin-up and spin-down electrons can tunnel into the well, while only the one that entered can tunnel out again, changing the equilibrium population to $1/(1+0.5 e^{\Delta E/k_BT})$. However, for a flat distribution of $\Delta  E_i$ the results below do not change after the integration over $\Delta E$.

\subsection{Summing the contribution from all TLFs}

 We start with the expression for the total charge induced on the island,
\begin{equation}
Q(t)=\sum _{i=1}^N \delta q_{i} \left[ f(\Delta E_i)-f(\Delta E_i+\delta E_i)\right] (1-e^{-t/\tau_{i}}) ,
\end{equation}

\noindent where the subscript \textit{i} denotes the individual values for the TLFs. We now assume that the TLFs are sufficiently numerous that we can replace the sum by an integral,
\begin{equation}
Q(t)=\int n(\vec{r},\vec{d},\Delta E,\tau) \delta q(\vec r, \vec d) \left[ f(\Delta E)-f(\Delta E+\delta E)\right] (1-e^{-t/\tau_{i}}) \ {\rm d}\Delta E {\rm d}\vec{r}\ {\rm d}\vec{d} \ {\rm d}\tau.
\end{equation}

\noindent We have introduced the density of TLFs, $n(\vec{r},\vec{d},\Delta E, \tau)$, which depends on location $\vec{r}$, displacement vector $\vec{d}$, TLF energy difference $\Delta E$, and switching time $\tau$.
We furthermore assume that the switching times $\tau_{i}$ have a logarithmic distribution between minimum and maximum switching times $\tau_{\rm min} = 1/\omega_{\rm max}$ and $\tau_{\rm max} = 1/\omega_{\rm min}$, independent of the distribution of $\vec d$ and $\Delta E$. This is consistent with the observed $1/f$ power spectrum for the noise \cite{MG_PRB} and would arise naturally from a roughly flat distribution of barrier heights in the case of thermal activation, and barrier widths in the case of quantum tunneling when the barrier height is the largest energy of the system. We can then perform the integral over $\tau$, and for times $\tau_{\rm min} \ll t \ll \tau_{\rm max}$ we find
\begin{equation}
Q(t)= \frac{\ln{(\omega_{\rm max} t)}+\gamma}{\ln{10}} \int n(\vec{r},\vec{d},\Delta E) \delta q(\vec{r},\vec{d}) \left[ f(\Delta E)-f(\Delta E+\delta E)\right] \ {\rm d}\Delta E {\rm d}\vec{r}\ {\rm d}\vec{d},
\end{equation}

\noindent where $\gamma$ is Euler's gamma. Further assuming that this density has a flat distribution with respect to the energy difference $\Delta E$, we can simplify this integral to

\begin{equation}
Q(t)=\frac{\ln{(\omega_{\rm max} t)}+\gamma}{\ln{10}} \int n(\vec{r},\vec{d})\ \delta q(\vec{r},\vec{d}) \delta E(\vec{r},\vec{d}) \ {\rm d}\vec{r}\ {\rm d}\vec{d},
\end{equation}

\noindent where $n(\vec{r},\vec{d})$ now denotes the TLF density per decade of frequency and per unit of energy difference. Using the expressions for $\delta q(\vec{r},\vec{d})$ (Eq.\ref{dq}) and $\delta E(\vec{r},\vec{d})$ (Eq.\ref{dE}), we arrive at
\begin{equation}
Q(t)= e^2 d^2 \Delta V \frac{\ln{(\omega_{\rm max} t)}+\gamma}{\ln{10}} \int n(\vec{r},\theta,\phi)\ e_G(\vec r)\ e_V(\vec r) \cos{\theta}  \left[\cos{\phi}\sin{\theta} \sin{\theta_V}+\cos{\theta}\cos{\theta _V}\right] \frac{\sin{\theta}}{4\pi}\ {\rm d}\vec{r}\ {\rm d}\theta\ {\rm d}\phi.
\end{equation}

We can now evaluate the integrals over TLF orientation angles $\theta$ and $\phi$ for the two different scenarios of a constant volume density and a constant surface density of TLFs. For a constant volume density $n_v$ of TLFs with random orientation, we arrive at:

\begin{equation}
\label{nVol}
Q_v(t)= \frac{1}{3} e^2 d^2 \Delta V n_v \frac{\ln{(\omega_{\rm max} t)}+\gamma}{\ln{10}}  \int_{V} \ e_G(\vec r)\ e_V(\vec r)\cos{\theta _V} {\rm d}\vec{r}.
\end{equation}

On the other hand, for a constant surface density $n_s$ of TLFs, with displacement vectors oriented normal to the surface, we arrive at:

\begin{equation}
\label{nSurf}
Q_s(t)= e^2 d^2 \Delta V n_s \frac{\ln{(\omega_{\rm max} t)}+\gamma}{\ln{10}} \int_{S} e_G(\vec r)\ e_V(\vec r)\cos{\theta _V} {\rm d}\vec{r}.
\end{equation}

\subsection{Temperature dependence of the step response}

In the case of {\em quantum tunneling} the effect of temperature is essentially to broaden the population according to the Fermi distribution. As shown above, this has no effect on the step response.

In the case of {\em thermal activation}, the switching time $\tau$ is determined by the barrier height $E_b$, the attempt frequency $\Omega$ and the temperature $T$ according to $\tau^{-1}=\Omega e^{-E_b/k_BT}$. Now consider a flat distribution of barrier heights, giving a density of $N$ TLFs per decade of switching time at temperature $T_0$. If we now double the temperature, all switching times decrease. If we consider temperature independent attempt rates as well as barrier heights, the TLF density increases to $2N$ TLFs per decade. Thus it is clear that in the case of thermal activation the TLF density per decade is proportional to temperature. Since the step response $Q(t)$ is proportional to the TLF density, it follows that the step response should also be proportional to temperature. Furthermore, if the induced energy shift is much larger than temperature, $\delta E \gg k_BT$, the voltage step will actually flatten out the TLFs which are thermally active at equilibrium, having a barrier height comparable to temperature, $E_b \sim k_B T$. These TLFs will thus switch immediately, and only those with a remaining barrier height $E_b'$ comparable to temperature will contribute to the slow charge drift. Assuming a flat distribution of final barrier heights $E_b'$, we find that with increasing step height $\Delta V$ the first set of TLFs that switch immediately will grow with $\Delta V$. On the contrary, the remaining set which contribute to the slow charge drift will actually be independent of $\Delta V$.

\section{II. FEM Calculation}

To determine the electrostatics of the joint system of the TLF ensemble and the SET, we implemented a numerical model using the sofware \textit{Comsol Multiphysics}. The geometry closely resembles that of the devices used in the experiments (Fig. S2). We take advantage of the fact that the SET is symmetric to use the computational resources efficiently. The substrate is taken to be silicon ($\epsilon_r = 12$) covered with 400 nm of oxide ($\epsilon_r = 4$). The SET and gate are built from a two-dimensional pattern, which is extruded by 70 nm along the normal to the substrate surface. The junctions are thus in the plane of the metal, and their thickness in the model is 5 nm. Above the substrate and the metal, a thick vacuum region ($\epsilon_r = 1$) is included in the model. The total size of the model is 6 $\mu$m by 2 $\mu$m by 4.4 $\mu$m, with symmetry conditions applied to the outer boundaries. Moving the boundaries further does not change the results appreciably.

In this geometry, we solve the Poisson equation in two different situations. For the first computation, we assign unit potential to all surfaces of the gate and zero potential to all exposed metal surfaces of the SET island and leads. This computation thus gives us the normalized electric field due to the gate voltage, $\vec e_G({\vec r})$, which changes the potential landscape of the TLFs.

In the second situation, we assign unit potential to the SET island and maintain zero potential at all other metal surfaces. The normalized virtual field $\vec e_V(\vec r)$ computed this way determines the influence that a TLF switching event has on the charge induced by the SET island [Eq.\,(\ref{dq})] for a TLF at any location within the model space.

\begin{figure}[h]
\includegraphics[width = 0.7\columnwidth]{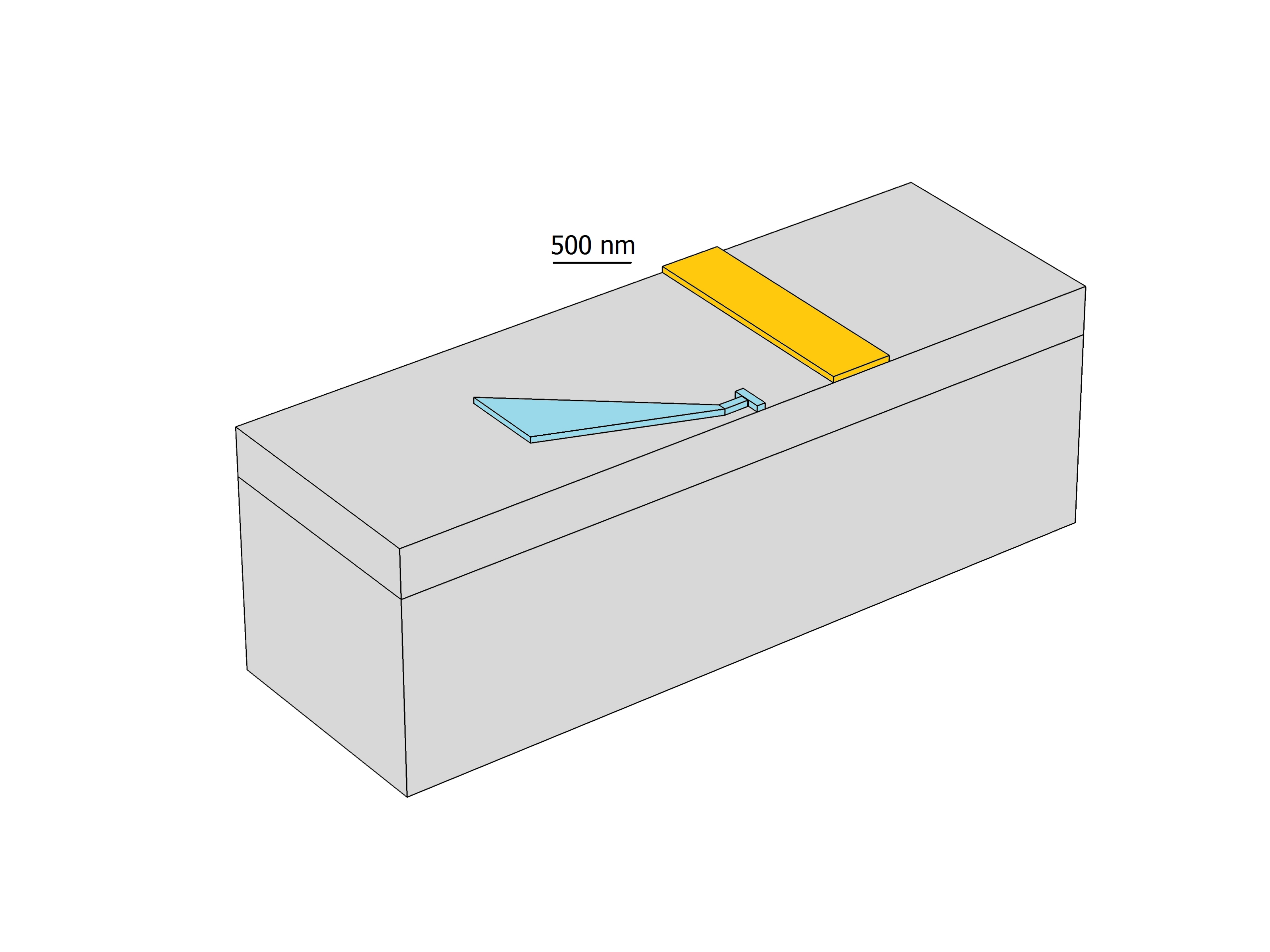}
\caption{Geometry of the model which was used for the numerical calculations. The yellow electrode is the gate and the cyan metals are the island and the leads. The numerical calculations are performed for half of the real geometry using the fact that the device is symmetric.}
\label{FigS2}
\end{figure}

We use the two computed fields to solve the integrals in Eqs.\,(\ref{nVol},\ref{nSurf}), which represent the case of bulk TLFs and surface TLFs, respectively. For the volume case, we calculate the volume integral $\int_V \vec e_V(\vec r) \cdot \vec e_G(\vec r) dV$ in the substrate and obtain a value of 478 nm. This value is used in Eq.\,(\ref{nVol}) to extract the volume density $n_v$. For the surface case, we calculate $\int_S \vec e_V(\vec r) \cdot \vec e_G(\vec r) dS$ over all exterior boundaries of the SET island, the leads and the gate (Table I). These values are used in Eq.\,(\ref{nSurf}) to calculate the surface density $n_s$. Based on physical considerations, both fields should be normal to the metal surfaces, and we verify that $\int_S \vec e_{V,\bot}(\vec r) \cdot \vec e_{G,\bot}(\vec r) dS$ does not differ from $\int_S \vec e_{V}(\vec r) \cdot \vec e_{G}(\vec r) dS$ due to numerical inaccuracy.

\begin{table}[h]
\renewcommand{\arraystretch}{1.5}
\caption{\label{tab:table1}%
Calculated values for the surface integral for different surfaces. }
\begin{tabular}{|l|c|}
\hline
Integrated surface (\textit{S})&$\int_S \vec e_{V}(\vec r) \cdot \vec e_{G}(\vec r)\,dS$ \\
\hline
island&-32.52 (nm) \\
leads&11.49 (nm) \\
gate&5.32 (nm) \\
\hline
\end{tabular}
\end{table}

We use a tetragonal mesh which is adapted to have very high density at the edges and boundaries of the metals. The mesh is further refined with a set of thin layers at these boundaries, to resolve accurately the fields that pertain to surface TLFs. The total number of mesh elements was 2844891.

To verify that the model represents the experimental devices accurately, we calculate the gate-island capacitance, by integrating the surface charge on the island with the gate at an elevated potential. The computed value, \\ Cg = 10aF, is in agreement with the experimental value.

We also compare the realistic model with a simplified version where the SET island and leads are represented as coaxial cylinders, with a gap between the end surfaces to represent the junctions. This geometry guarantees that numerical errors are not introduced by electric field concentration at the corners of the SET metal. The deviations between this model and the actual structure are quite small, indicating that the numerical solutions are reliable.

\bibstyle{apsrev}
\bibliography{References}